\documentclass[prl,aps,twocolumn,showpacs]{revtex4}

\usepackage{graphicx}

\begin{document}

  \title{Density Functional Theory Characterization of the Multiferroicity \\
in Spin Spiral Chain Cuprates}

  \author{H. J. Xiang}
  \affiliation{Department of Chemistry, North Carolina State University, Raleigh,
    North Carolina 27695-8204}

  \author{M.-H. Whangbo}
  \thanks{Corresponding author. E-mail: mike\_whangbo@ncsu.edu}

  \affiliation{Department of Chemistry, North Carolina State University, Raleigh,
    North Carolina 27695-8204}

\date{\today}

\begin{abstract}
The ferroelectricity of the spiral magnets LiCu$_2$O$_2$ and LiCuVO$_4$
was examined by calculating the electric polarizations of their spin
spiral states on the basis of density functional theory with
spin-orbit coupling. Our work unambiguously reveals that spin-orbit coupling is
responsible for the ferroelectricity with the primary contribution
from the spin-orbit coupling on
the Cu sites, but the asymmetric density distribution
responsible for the electric polarization occurs mainly around the O
atoms. The electric polarization is calculated to be much greater for
the $ab$- than for the $bc$-plane spin spiral. The observed
spin-spiral plane is found to be consistent with the observed 
direction of the electric polarization for LiCuVO$_4$, but inconsistent
for LiCu$_2$O$_2$.
\end{abstract}

\pacs{75.80.+q,77.80.-e,71.15.Rf,71.20.-b}
%75.80.+q Magnetomechanical and magnetoelectric effects, magnetostriction
%77.80.-e Ferroelectricity and antiferroelectricity
%71.15.Rf Relativistic effects
%71.20.-b Electron density of states and band structure of crystalline solids

\maketitle
In the past several years there has been a revival of interest in
understanding magnetic ferroelectrics because of their potential
applications in novel magnetoelectric and magneto-optical devices
\cite{Kimura2003, Hur2004, Ikeda2005, Xiang2007, Lawes2005}. In these
multiferroics, the paramagnetic phase is centrosymmetric, and
electrical polarization appears only at the transition to a
magnetically ordered phase, which is responsible for removing the inversion symmetry thereby
generating a polar field. Two mechanisms of multiferroicity have been
proposed in the literature. The exchange striction due to the
symmetric parts of the exchange coupling appears relevant for the
multiferroicity in systems such as RMn$_2$O$_5$ \cite{Chapon2006}. The
ferroelectricity (FE) in spiral magnets has been suggested to originate from
spin-orbit coupling (SOC) \cite{Katsura2005, Sergienko2006,
  Mostovoy2006, Jia2006}. However, this suggestion has never been
verified by first principles electronic structure studies.

Very recently, FE was
discovered in two cuprates LiCu$_2$O$_2$ \cite{Park2007} and LiCuVO$_4$
\cite{Naito2007}. These oxides contain
spin-frustrated CuO$_2$ ribbon chains that are made up of edge-sharing
CuO$_4$ squares with spin-$\frac{1}{2}$ magnetic ions Cu$^{2+}$, for which an
unpaired spin resides in the $e_g$ orbital (commonly referred to as
the $d_{x^2-y^2}$ orbital, but the $d_{xy}$ orbital in the local axis system
adopted in the present work). 
Due to the competition between
nearest-neighbor (NN) ferromagnetic and next-nearest-neighbor
antiferromagnetic interactions in each CuO$_2$ chain, a spin spiral state
can set in along the chain direction (i.e., the $b$ direction) at low
temperature \cite{Bursill1995}. Neutron diffraction studies showed
that the Cu$^{2+}$ moments lie in the CuO$_2$ ribbon plane (i.e., the
$ab$-plane) in the spin spiral state of both LiCu$_2$O$_2$ \cite{Masuda2004}
and LiCuVO$_4$ \cite{Gibson2004}. Results of a recent ESR study \cite{Mihaly2006} on
LiCu$_2$O$_2$ are consistent with the $ab$-plane spin spiral.
Below the temperature of the
spiral-magnetic order, the electric polarization is found to be along the
$a$ direction in LiCuVO$_4$ \cite{Naito2007}, but along the $c$ direction
in LiCu$_2$O$_2$ \cite{Park2007}. 
This difference in the electric polarization directions
is puzzling because, according to the Katsura-Nagaosa-Balatsky
(KNB) model \cite{Katsura2005}, the electric polarization along the
$a$-direction is associated with the $ab$-plane spin spiral, and that
along the $c$-direction with the $bc$-plane spin spiral. However, it is
unclear whether the KNB model is applicable to the $e_g$ systems \cite{Naito2007}
because it was formulated for a $t_{2g}$ system.
Furthermore, in their recent study of the spiral magnet LiCuVO$_4$, Jia
{\it et al.} reported \cite{Jia2007} that the SOC on the ligand O 2p
orbitals plays a more important role than does the SOC on the Cu 3d
orbitals. It is important to verify whether or not this conclusion is
correct.

\begin{table}
  \caption{Electric polarizations (in units of $\mu C/m^2$) of LiCu$_2$O$_2$ and LiCuVO$_4$ calculated
  for the $ab$-, $ac$-, and $bc$-plane spin spiral states of
  Fig.~\ref{fig2} using the experimental (EXP) centrosymmetric
  structures and the optimized (OPT) structures. } 
  \begin{tabular}{ccccccc}
    \hline
    \hline
    & $\mathbf{P}(ab)$  & $\mathbf{P}(ac)$  & $\mathbf{P}(bc)$  \\
    \hline
    LiCu$_2$O$_2$ (EXP) & (51.4,0,0) & (0,0,0)& (0,0,-9.8) \\
    LiCu$_2$O$_2$ (OPT) & (441.4,0,0) & & (0,0,-191.8) \\
    LiCuVO$_4$ (EXP) & (103.5,0,0) &(0,0,0) &(0,0,-15.7) \\
    LiCu2O4 (OPT) & (595.6,0,0) & & (0,0,-223.0)\\
    \hline
    \hline
  \end{tabular}
  \label{table1}
\end{table}

%\clearpage

\begin{figure}
  \includegraphics[width=6.0cm]{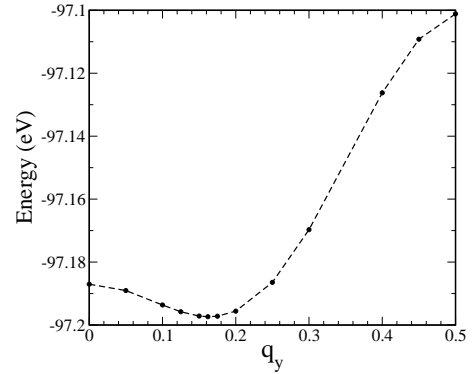}
  \caption{Total energy E($\mathbf{q}$) calculated for the spin spiral
    state of LiCu$_2$O$_2$ as a function of $\mathbf{q}=(0.5,q_y,0)$
    on the basis of the non-collinear LDA$+$U method using a unit cell
    containing four formula units. }
  \label{fig1}
\end{figure}

%\clearpage

\begin{figure}
  \includegraphics[width=6.0cm]{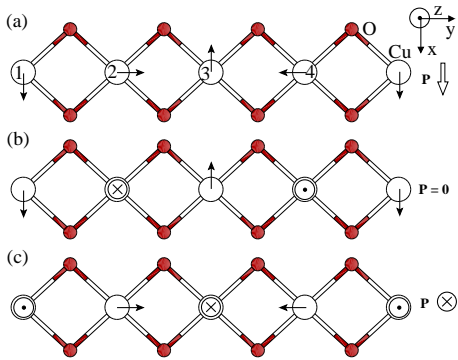}
  \caption{(color online) 
Spin arrangements in the (a) $ab$-, (b) $ac$-, and (c) $bc$-plane spin
spiral states using the $1\times4\times1$ supercell. The large
white and the small red circles represent the Cu and O atoms,
respectively. The legends $\odot$ and $\otimes$ denote z and $-$z
directions, respectively. The directions of the electric polarization
corresponding to a given spiral plane are also shown. In the present
work, the $a$, $b$ and $c$ axes correspond to the local $x$, $y$ and
$z$ axes, respectively.}
  \label{fig2}
\end{figure}

%\clearpage

In this Letter, we address the aforementioned issues on the
basis of 
non-collinear fully relativistic density functional calculations for
LiCu$_2$O$_2$ and LiCuVO$_4$. The electric polarizations associated with the
non-collinear spin-orbit coupled states are calculated using the
modern polarization theory \cite{King-Smith1993} for the first
time. To the best of our knowledge, this is the first study on
multiferroicity in spiral magnets using first principles method with
the SOC effect included.
Our electronic structure calculations were carried out using the local
density approximation (LDA) on the basis of the projector augmented
wave method \cite{PAW} encoded in the Vienna ab initio simulation
package (VASP) \cite{VASP}. The plane-wave cutoff energy was set to 400
eV. To properly describe the strong electron correlation in the 3d
transition-metal oxide, the LDA plus on-site repulsion U method
(LDA$+$U) was employed \cite{Liechtenstein1995}. In the following, we
report results obtained with U $=$ 6 eV and J $=$ 0 eV on Cu, but
our results with other U values between $4-7$ eV are similar.

In the absence of SOC, the incommensurate spiral magnetic order can be
simulated without resorting to the supercell technique due to the
generalized Bloch theorem \cite{Sandratskii1998}. A neutron
diffraction study \cite{Masuda2004} showed that the spin structure of
LiCu$_2$O$_2$ is helimagnetic with propagation vector
$\mathbf{q}=(0.5,0.174,0)$. We performed a series of calculations
\cite{NC} for the spin spiral state with propagation vector
$\mathbf{q}=(0.5,q_y,0)$. The dependence of the electronic energy E($\mathbf{q}$)
upon $q_y$ is presented in Fig.~\ref{fig1}, which shows that the
energy at $q_y=0$ (i.e., the FM arrangement between all NN's) is lower
in energy than that at $q_y=0.5$ (i.e., the AFM arrangement between
all NN's). This is in accord with the FM nature of the NN exchange
interaction. The energy minimum occurs at around $q_y=0.165$, in good
agreement with the experimental value of $q_y=0.174$. This agreement
indicates that the spin spiral ground state of a system with spin
frustrated CuO$_2$ ribbon chains can be predicted by non-collinear
LDA$+$U calculations. 

To see if the LDA$+$U method can describe the FE in
LiCu$_2$O$_2$ as well, we calculated the electric polarization of the spin
spiral state with $q_y=0.174$ using the Berry phase method
\cite{King-Smith1993}. This calculation leads to negligible electric
polarization. As a possible reason for this failure, we considered the
lack of geometry relaxation in the spin spiral ground
state. 
However,
LDA$+$U calculations for the fully optimized structure of LiCu$_2$O$_2$ does not
lead to any appreciable electric polarization. Similarly, our LDA$+$U
calculations for the spin spiral state of LiCuVO$_4$ show no electric
polarization. Consequently, the exchange striction is not an
appropriate mechanism of electric polarization for LiCu$_2$O$_2$ and
LiCuVO$_4$. 

The above observation led us to examine SOC effects on the electric
polarization in the spin spiral states of LiCu$_2$O$_2$ and LiCuVO$_4$. We
carry out LDA+U+SOC calculations for the $\mathbf{q}=(0,0.25,0)$ spiral
states \cite{SOC} with three different spin spiral arrangements, shown
in Fig.~\ref{fig2}, by using a $1\times4\times1$ supercell. Our
calculations for LiCu$_2$O$_2$ give rise to substantial electric
polarizations (see Table~\ref{table1}). The electric polarization
$\mathbf{P}$ is calculated to be along the $a$-direction for the
$ab$-plane spin spiral, and along the $-c$-direction for the
$bc$-plane spin spiral \cite{footnote1}. When the spin spiral is in the $ac$-plane, the
electric polarization is calculated to be zero. The same results are
also obtained for LiCuVO$_4$. Thus, for both LiCu$_2$O$_2$ and LiCuVO$_4$, the
directions of the calculated electric polarizations are consistent
with the prediction of the KNB model. This agreement is somewhat
surprising since the KNB model was derived for a linear trimer M-O-M
with $t_{2g}$  transition  metal ions M \cite{Katsura2005}.
The calculated electric polarizations are anisotropic, e.g., the electric
polarizations of LiCu$_2$O$_2$ are 51.4 and 9.8 $\mu C/m^2$ for the $ab$-
and $bc$-plane spin spiral arrangements, respectively. Note that the
anisotropy in electric polarizations is not predicted by the KNB
model.

Our calculations of electric polarizations show that the observed
spin-spiral plane is consistent with the observed direction of the
electric polarization for LiCuVO$_4$, but this is not the case for
LiCu$_2$O$_2$. It is of interest to consider a probable reason for the
latter discrepancy.  For LiCuVO$_4$ \cite{Buttgen2007}, the planes of their spin spiral can be flipped by the
application of an external magnetic field. For LiCu$_2$O$_2$
\cite{Park2007}, the direction of the electric polarization can be
flipped by an external magnetic field. 
In the ideal crystal structure of
LiCu$_2$O$_2$, the CuO$_2$ ribbon chains with Cu$^{2+}$ ions are interconnected by
linear O-Cu-O bridges with diamagnetic Cu$^+$ ions. Matsuda {\it et al.}
reported that the actual composition of their ``LiCu$_2$O$_2$'' sample is
given by Li$_{1.16}$Cu$_{1.84}$O$_{2.01}$, in which 16\% Cu$^{2+}$ sites of the CuO$_2$
ribbon chains are replaced with diamagnetic Li$^+$ ions 
due to a good match of the ionic radii of Li$^+$  and Cu$^{2+}$
while 16\% Cu$^+$
ions in the linear O-Cu-O bridges become Cu$^{2+}$ ions due to the charge
balance requirement. This implies that the spins of the Cu$^{2+}$ ions
residing in between the CuO$_2$ ribbon chains can act as a source of
``external''  magnetic field for the CuO$_2$ ribbon chains, and hence can
influence the plane of spin spiral in the CuO$_2$ ribbon chains. 
Similarly, the Ising behavior of CuFeO$_2$ with high spin Fe$^{3+}$ (d$^5$) ions
below 14 K has been explained in terms of a uniaxial magnetic field
associated with oxygen defects \cite{whangbo}.
Since the nonstoichiometry of ``LiCu$_2$O$_2$'' samples would depend strongly on
synthetic conditions, it is highly  desirable to measure the plane of spin
spiral and the direction of electric polarization by using the same
sample. 

To test the conclusion of Jia {\it et al.} \cite{Jia2007} that the SOC
on the O 2p orbitals is more important for the ferroelectric
polarization of LiCuVO$_4$ than is the SOC on the Cu 3d orbitals, we
carried out LDA+U+SOC calculations for LiCuVO$_4$ with the $ab$-plane
spin spiral state by switching off the SOC on either the Cu or the O
sites. When the SOC is neglected on the Cu sites but kept on the O
sites, the electric polarization is calculated to be along the $-a$
axis with $P_a=-29.8$ $\mu C/m^2$. If the SOC is kept on the Cu sites
but neglected on the O sites, the electric polarization is calculated
to be along the $a$ axis with $P_a=127.1$ $\mu C/m^2$. The sum of the
above two polarizations is very close to the electric polarization
($P_a=103.5$ $\mu C/m^2$) calculated when the SOC is kept on both Cu and
O sites. Therefore, the primary contribution to the electric
polarization comes from the SOC on the Cu sites, which is greater than
that from the SOC on the O sites by a factor of approximately four. 

\begin{figure}
  \includegraphics[width=7.0cm]{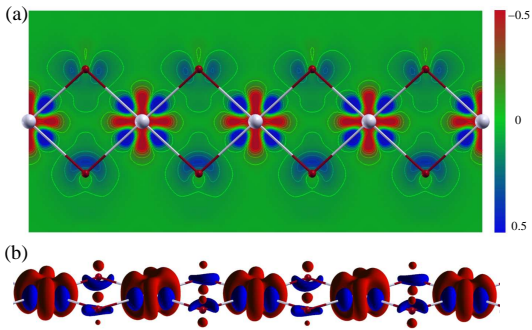}
  \caption{(color online)
    (a) Cross section view of the electron density difference between the
    LDA$+$U$+$SOC and LDA$+$U calculations for the $ab$-plane
    spin spiral state (Fig.~\ref{fig2}(a)) of LiCuVO$_4$. (b)
    Perspective view of an isosurface calculated for
    the electron density difference between the LDA$+$U$+$SOC and
    LDA$+$U results for the $bc$-plane spin spiral state
    (Fig.~\ref{fig2}(c)) of LiCuVO$_4$. The red and blue surfaces
    represent $-0.25$ and $0.25$, respectively.}
  \label{fig3}
\end{figure}

%\clearpage

\begin{figure}
  \includegraphics[width=8.5cm]{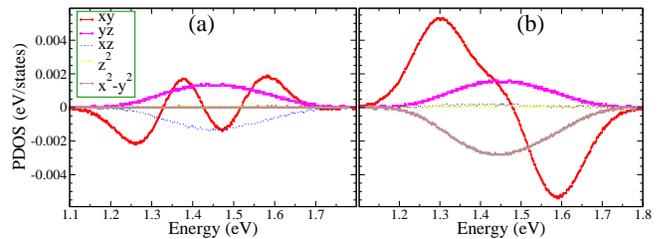}
  \caption{(color online)
    Difference in the PDOS plots of the adjacent Cu atoms, i.e., the
    Cu2 and Cu1 atoms of  Fig.~\ref{fig2}(a), in the (a) $ab$- and (b)
    $bc$-plane spin spiral state of LiCuVO$_4$. The energy region
    plotted corresponds to the hole $xy$ state of LiCuVO$_4$.}
  \label{fig4}
\end{figure}

The dominance of the SOC on the Cu sites does not necessarily mean
that the asymmetric charge distribution, needed for nonzero electric
polarization, is associated mainly with the Cu atoms. To examine how
the FE arises from the spiral magnetic state, we analyze
the electron density distribution of the spin spiral state of LiCuVO$_4$
by plotting the difference between the electron density of the
LDA$+$U$+$SOC calculation and that of the LDA$+$U calculation. For the
case of the $ab$-plane spin spiral, Fig.~\ref{fig3}(a) shows that the
difference density around each Cu ion is almost symmetric, and hence
contributes little to the electric polarization. 
However, an asymmetric difference density is found on the O atoms such
that a nonzero electric polarization sets in along the $a$-direction. 
Fig.~\ref{fig3}(b) shows a perspective view of
the difference density plot for the case of the $bc$-plane spin
spiral. Again, the difference density around each Cu atom is nearly
symmetric, and an asymmetric difference density is found at each O
atom largely along the $c$-direction eventually leading to the electric
polarization along the $c$-direction. 

From the viewpoint of local bonding, 
the SOC at each Cu site mixes into the hole $xy$ state other
3d-states (i.e., the $x^2-y^2$, $xz$, and $yz$ states). The extent of the 
mixing for each 3d-state 
depends on the spin direction at a given Cu site because of the
SOC term $\lambda S \cdot L$. This gives rise
eventually to the asymmetric electron density distribution on the O
sites because each 3d state of a given Cu site has the 2p orbitals of
its surrounding O atoms combined out-of-phase with the Cu 3d
orbital. 
We confirmed this point in two ways. First, 
we analyzed the partial density of states (PDOS)
for the hole states of LiCuVO$_4$ (i.e., the unoccupied band associated
with the down-spin $xy$ state from each Cu site), which are located in
the energy range from 1.1 to 1.8 eV above the Fermi level. In the spin
spiral state, the neighboring Cu ions in each CuO$_2$ chain with
different spin directions will have slightly different 3d PDOS because the mixing between the
3d-states induced by the SOC depends on the spin direction. As
expected, the Cu1 and Cu2 ions [as defined in Fig.~\ref{fig2}(a)] show
differences in their PDOS not only for the $ab$-plane spin spiral
(Fig.~\ref{fig4}(a)) but also for the $bc$-plane spin spiral
(Fig.~\ref{fig4}(b)). 
Second, we extended the tight-binding calculations of Jia {\it et al.}
\cite {Jia2006} to a tetramer model in which two Cu ions are bridged
by two O ions as found in the CuO$_2$ ribbon chains, and carried out
numerical solutions by using all 3d orbitals of
Cu and all 2p orbitals of O. As expected, the occupations of
the $x^2-y^2$, $xz$, and $yz$ states at a given Cu site are found to depend on
the spin direction of that site. In addition to a transverse electric
dipole moment, a nonzero polarization is found along the Cu chain
direction. This latter longitudinal component oscillates as the spin
spiral progresses along the chain, and hence its effect is canceled
out. Consequently, the overall polarization is calculated to be along
the $a$ and $c$ directions in the case of the $ab$- and $bc$-plane
spin spiral arrangements, respectively.

Concerning the effect of SOC on the FE in spiral magnets,
there is a debate as to whether
the FE arises from pure electronic effects
\cite{Katsura2005,Jia2007} or displacements of ions
\cite{Sergienko2006}. Using first principles calculations without
taking SOC into account, Picozzi {\it et al.} \cite{Picozzi2007}
showed that both factors are active for the FE in HoMnO$_3$. In our
calculations described above, we used the 
experimental centrosymmetric structure for both LiCu$_2$O$_2$
\cite{Berger1992} and LiCuVO$_4$ \cite{Lafontaine1989}. To evaluate the
effect of polar atomic displacements on the electric polarization, we
optimized the crystal structures by LDA+U+SOC calculations until the
atomic forces become less than 0.01 eV/\AA \cite{footnote2}. The calculated electric
polarizations using the resulting relaxed structures are shown in
Table~\ref{table1}, which shows that the direction of the electric
polarizations remains unchanged but the magnitude of the electric
polarization becomes strongly enhanced by the structural
relaxations. It is found that the Cu$^{2+}$ ions move along the $-a$ and
$c$ directions in the case of the $ab$- (Fig.~\ref{fig2}(a)) and
$bc$-plane (Fig.~\ref{fig2}(c)) spin spiral states, respectively. With
respect to the centrosymmetric structure, however, the largest
displacement of the Cu$^{2+}$ ions is about 0.0004 \AA, which are hardly
detectable in experiment. Note that the experimental electric
polarization at 2.5 K for LiCu$_2$O$_2$ is 4 $\mu C/m^2$, which is
reasonably well reproduced by the calculation using the experimental
structure (i.e., 9.8 $\mu C /m^2$). The electric polarization 191.8
$\mu C/m^2$ calculated by using the relaxed structure is much
greater. Since the polar atomic displacements associated with the
relaxation are extremely small, it is probable that their effects are
negated by zero point vibrational effects.  

To probe how one might improve the effect of multiferroicity, it is
necessary to examine how SOC affects electric polarization. For this
purpose, we performed LDA+U+SOC calculations for LiCuVO4 by treating
the speed of light $c$ as an adjustable parameter and noting that the
strength of SOC is inversely proportional to the square of the speed
of light $c$ \cite{Kunes2001}. These hypothetical calculations show that the
magnitude of electric polarization is proportional to the strength of
SOC. Since the SOC of an atom increases in strength with increasing
the atomic number, high-temperature multiferroics with large electric
polarization are expected from spin-spiral systems with 4d or 5d
transition metal elements. Another important factor is that the
critical temperature of the spin spiral ordering should be high.

In summary, our study of LiCuVO$_4$ and LiCu$_2$O$_2$ reveals that SOC is responsible for their
FE, the primary contribution to the FE
arises from the SOC on the Cu sites, but the asymmetric density
distribution responsible for the electric polarization occurs mainly
around the O atoms. For both LiCuVO$_4$ and LiCu$_2$O$_2$, the electric
polarization is calculated to be much greater for the $ab$- than for
the $bc$-plane spin spiral.  The observed spin-spiral plane is
consistent with the observed direction of the electric polarization
for LiCuVO$_4$. However, this is not the case for LiCu$_2$O$_2$. To resolve the
latter discrepancy, further experimental and theoretical studies are
necessary.

%%Acknowledgements
The research was supported by the Office of Basic Energy Sciences, Division of
Materials Sciences, U. S. Department of Energy, under Grant No. DE-FG02-86ER45259. 
We thank Erjun Kan for useful discussions.

%\clearpage

%\clearpage


\begin{thebibliography}{99}
\bibitem{NC}Here, a larger plane-wave cutoff energy, i.e. 500 eV, is
  adopted for better convergence.
\bibitem{SOC}The electric
  polarization can be understood as a sum of the contributions from
  individual  CuO$_2$ chains. The sign of the contribution from each
  CuO$_2$ chain depends  largely on its spin chirality.
  This supports the simulation of a spiral
  state with $\mathbf{q}=(0,0.25,0)$ instead of
  $\mathbf{q}=(0.5,0.25,0)$, which requires a $2\times4\times1$
  supercell.
  Test calculations using $1\times6\times1$ supercell with
  the corresponding $\mathbf{q}=(0,1/6,0)$ give qualitatively similar
  results.
\bibitem{footnote1} Our LDA+U+SOC calculations show spontaneous
  electric polarization in the spin spiral state even when the atom
  positions are kept to have the centrosymmetric space group. This is
  so because the mixing between electronic states induced by SOC leads
  to an electronic state that lacks inversion symmetry.
\bibitem{footnote2} In our structural optimizations, the P1 symmetry
  was used, i.e., no symmetry constraint was imposed. Obviously,
  inversion symmetry is lost after geometry optimization. Within the
  tolerance limit of 0.00002 \AA, the space groups of the ab- and
  bc-plane spin spiral states of LiCu$_2$O$_2$ are Pc and P21, respectively,
  while those of the ab- and bc-plane spin spiral states of LiCuVO$_4$
  are both P1.
\end{thebibliography}
\end{document}